\documentclass[onecolumn,prl,superscriptaddress,showpacs,10pt,aps,notitlepage]{revtex4-1}
\pdfoutput=1

\newif\ifarxiv\arxivtrue

\usepackage{amsmath}
\usepackage{amssymb}
\usepackage{graphicx}
\usepackage{color}

\usepackage[T1]{fontenc}
\usepackage[latin9]{inputenc}

\usepackage{amsthm}
\usepackage{bbm}
\usepackage{mathtools}

\ifarxiv\fi

\def\lettersection#1{\ifarxiv\section{#1}\else\par\vskip12pt\noindent{\it #1.\/\ }\fi}

\def\explorer{\ifarxiv\cite{explorerA}\else\cite{explorerP}\fi}

\usepackage{tikz}
\usetikzlibrary{arrows,matrix}
\usetikzlibrary{decorations.pathreplacing}
\usetikzlibrary{decorations.markings}
\usetikzlibrary{shapes,arrows,positioning,patterns}

\def\ya{-.02}
\def\yb{-.22}

\def\coloronline{\ifarxiv\else(Color online) \fi}

\def\tr{\operatorname{tr}}
\def\idty{{\leavevmode\rm 1\mkern -5.4mu I}} 

\def\Ir{{\mathbb Z}}

\def\norm #1{\Vert #1\Vert}

\def\mod{{\mathop{\rm mod}\nolimits}}

\def\tr{\mathop{\rm tr}\nolimits}
\def\abs#1{\vert#1\vert}

\def\matrix#1{{\left(\begin{array}#1\end{array}\right)}}

\def\im{\Im m}

{{\def\HH{{\mathcal H}}

\def\espace{{\mathcal E}}

\def\ig{\mathbf I} 
\def\six{\mathop{\rm si}\nolimits}
\def\sixR{\protect\overrightarrow{\six}}
\def\sixL{\protect\overleftarrow{\six}}

\ifarxiv\def\dg{^*}\else\def\dg{^\dagger}\fi

\def\ph{\eta}\def\rv{\tau}\def\ch{\gamma}

\definecolor{ForestGreen}{RGB}{34,139,34}

\usepackage{hyperref}

\begin{document}
\title{Bulk-edge correspondence of one-dimensional quantum walks
}

\author{C. Cedzich}
\affiliation{Institut f\"ur Theoretische Physik, Leibniz Universit\"at Hannover, Appelstr. 2, 30167 Hannover, Germany}
\author{F.~A. Gr\"unbaum}
\affiliation{Department of Mathematics, University of California, Berkeley CA 94720}
\author{C. Stahl}
\affiliation{Institut f\"ur Theoretische Physik, Leibniz Universit\"at Hannover, Appelstr. 2, 30167 Hannover, Germany}
\author{L. Vel\'azquez}
\affiliation{Departamento de Matem\'{a}tica Aplicada \& IUMA,  Universidad de Zaragoza,  Mar\'{\i}a de Luna 3, 50018 Zaragoza, Spain}
\author{A.~H. Werner}
\affiliation{Dahlem Center for Complex Quantum Systems, Freie Universit\"at Berlin, 14195 Berlin, Germany}
\author{R.~F. Werner}
\affiliation{Institut f\"ur Theoretische Physik, Leibniz Universit\"at Hannover, Appelstr. 2, 30167 Hannover, Germany}

\begin{abstract}
  We outline a theory of symmetry protected topological phases of one-dimensional quantum walks. We assume spectral gaps around the symmetry-distinguished points +1 and -1, in which only discrete eigenvalues are allowed. The phase classification by integer or binary indices extends the classification known for translation invariant systems in terms of their band structure. However, our theory requires no translation invariance whatsoever, and the indices we define in this general setting are invariant under arbitrary symmetric local perturbations, even those that cannot be continuously contracted to the identity. More precisely we define two indices for every walk, characterizing the behavior far to the right and far to the left, respectively. Their sum is a lower bound on the number of eigenstates at +1 and -1. For a translation invariant system the indices add up to zero, so one of them already characterizes the phase. By joining two bulk phases with different indices we get a walk in which the right and left indices no longer cancel, so the theory predicts bound states at +1 or -1. This is a rigorous statement of bulk-edge correspondence. The results also apply to the Hamiltonian case with a single gap at zero.
\end{abstract}

\pacs{
05.60.Gg  
03.65.Db  
03.65.Vf  
}

\maketitle

\ifarxiv
{}
\else
\fi  

\ifarxiv \parskip=12pt
\section{Introduction}\fi

Topological phases play an important role in the classification of quantum matter, e.g. in the distinction between topological and ordinary insulators in lattice systems \cite{KaneMeleQSH,KaneMeleTopOrder}. Such phases are intrinsic properties of physical systems exhibiting (a set of) symmetries (for a review, see \cite{HasanKaneReview} and references therein). A key feature of topological insulators is, that in contrast to ordinary phase transitions covered by the so-called Landau theory \cite{Landau}, the symmetries of the system remain unbroken during transitions between distinct topological phases.

Over the past years, such phenomena gained a lot of attraction both in theory and experiments due to their robustness against local perturbations which opens a wide range of applications from spintronics to topological quantum computation \cite{KitaevLectureNotes,KitaevModel,KitaevAnyons,AnyonsUniversalQuantumComp}.

A key intuition about topological phases is a principle known as the bulk-edge correspondence. Loosely speaking it states that if two systems in distinct phases are joined, a bound state should emerge near the interface. Moreover, this should be true irrespective of how the systems are joined. This additional stability distinguishes topologically protected bound states from a much more commonplace phenomenon: Whenever a system is perturbed locally, for example by introducing an impurity or defect, we know that bound states might appear \cite{Ashcroft}. Hence even if two systems in the same topological phase are joined, bound states will typically arise, but these will not be topologically stable and can be eliminated by engineering the crossover differently.

Classification of topological phases is well understood for translation invariant bulk systems. In that case the topological properties can be stated in terms of the winding properties of the energy bands over the quasi-momentum space, technically expressed by the K-theory of vector bundles \cite{kitaevPeriodic}. However, this structure is lost as soon as a phase boundary is introduced. Hence, as remarked in \cite{Kita2}, any theory of bulk-edge correspondence has to deal with non-translation invariant systems from the outset. It also has to deal with a vastly larger set of perturbations: In the translation invariant case, finitely many parameters suffice to characterize all systems with given maximal step size, but without translation invariance we obviously have to deal with infinitely many perturbation parameters.  In addition to small global perturbations,  we can have local perturbations, which cannot be achieved in many small steps, and are hence not amenable to standard deformation arguments \cite[Sect.~IV.E]{Kita2}. It is not a priori clear that a meaningful classification stable under all such perturbations exists. But, as we show in this paper, the combination of a gap condition with local symmetry sufficiently tames the possible behaviours at infinity to allow such a phase classification. It even turns out that this classification in the general case is {\it the same} as for the bulk: Indeed, every class in our classification can be realized by joining two translation invariant systems, and the pair of their bulk indices characterizes the class. This is the strongest statement of bulk-edge correspondence one could hope for.

The aim of our paper is to describe the basics of a rigorous theory of bulk-edge correspondence in this sense, and to sketch the key ideas of a proof, of which the details will appear elsewhere \cite{LongVersion}. The setting chosen is that of discrete-time evolutions of single-particles with internal degrees of freedom, so-called ``quantum walks'' \cite{Ambainis2001,Grimmet,SpaceTimeCoinFlux,TRcoin,GenMeasuringDevice,electric}. As such, quantum walks have recently attracted much attention as a computational resource \cite{AmbainisKR05,Magniez:2011ke,Anonymous:rSnqW2vc,Lovett:2010ff,SearchOnFractalLattice,Quenching}. In particular quantum walks have been shown to exhibit a rich variety of quantum effects such as Landau-Zener tunneling \cite{landauzehner}, the Klein paradox \cite{kurzwel:2008dm}, Bloch oscillations \cite{Gensketal}. By taking into account on-site interactions between two particles performing a quantum walk, the formation of molecules has been established \cite{molecules}. Local defects usually generate bound states \cite{Luis}. Eigenstates can also be generated by i.i.d.\ random \cite{dynloc,dynlocalain} and quasi-periodic \cite{Shikano,fillman2015spectral} coin choice leading to Anderson localization or by suitably modified rules to give infinitely degenerate subspaces without transport (Konno localization \cite{ThreeKonnos,LoKonno}). 

Quantum walks have been experimentally realized in such diverse physical systems as neutral atoms in optical lattices \cite{Karski:2009}, trapped ions \cite{Zaehringer:2010bs,Schmitz2009}, wave guide lattices \cite{Peruzzo:2010co,PhysRevLett.108.010502} and light pulses in optical fibres \cite{Schreiber:2010cl,Schreiber:2012ix} as well as single photons in free space \cite{PhysRevLett.104.153602}. 

The quantum walks considered in this manuscript obey a suitable subset of discrete symmetries $\{\ph,\rv,\ch\}$, commonly referred to as ``particle-hole'', ``time-reversal'' and ``chiral'' symmetry. It is well known from the theory of bulk phases that in one dimension such symmetries must be imposed on the system as well as all its perturbations \cite{MPSphaseII,MPSphaseI}. In accordance with this theory we select the subset of the symmetries of the tenfold way \cite{Altland-Zirnbauer} for which non-trivial phases are expected. In contrast to the 'periodic table' in \cite{kitaevPeriodic} we define and compute the invariants for the various symmetry types in an elementary group theoretical way.

Remarkably, our results also apply to the continuous time, i.e., (static) Hamiltonian case. However, we chose chiefly the discrete time setting, which applies to periodically driven ``Floquet'' systems,  because, on one hand, we plan such an experimental realization in an optical lattice with neutral atoms \cite{Bonn}, and on the other the discrete time case has some additional intricacies: First, there are two symmetry-related spectral gaps, at  $\pm1$ for the unitary transition operator, instead of the single gap at $0$ in the Hamiltonian case, and secondly there are local perturbations which cannot be contracted to the identity while preserving the symmetry. In order to achieve a classification, which is invariant also under non-contractible local perturbations, new methods beyond the usual deformation arguments are required.

An important input for our work were the publications of Kitagawa et al. \cite{Kita,Kita2} and subsequent work by Asb{\'o}th \cite{Asbo1,Asbo2}. These papers give examples which realize many possible bulk symmetry classes. Supporting evidence for the bulk-edge principle is given by numerically finding the dynamical signature of bound states in joined systems and observing that they appear exactly in situations in which the bulk-edge principle predicts them \cite{Kita,Kita2,Asbo1,ChandrashekarTopEntanglement}.

In this note we prove this observation in full generality. Naturally, with the benefit of a precise mathematical formulation, we can also sharpen some of the claims and statements in the aforementioned papers. We caution the reader, however, that our proof so far covers only 4 of the 5 symmetry types considered in \cite{Kita}, and is confined to the 1D lattice case.

\begin{figure}
	\begin{tikzpicture}[
	scale=1.5,
	font=\footnotesize,
	cont/.style={line width=3,green!80!black}
	]
	\def\hamx{2.8}
	\tikzset{
		rot/.style={rotate around={0:(\hamx,0)}}
	}
	\def\thsq{.08}
	\def\thp{.03}
	\colorlet{colorp}{red}		
	\colorlet{colorgap}{blue!80!black}
	\draw ({-1.0-5*\thsq},1) node(a) {a)};
	\draw ({\hamx-1.0},1) node(b) {b)};

	\draw[gray,rot] (\hamx,-1)  -- (\hamx,1);
	\draw[cont,rot] (\hamx,-1)  -- (\hamx,-.5);
	\draw[cont,rot] (\hamx,1)  -- (\hamx,.5);
	\foreach \y in {.4,.25,-.25,-.4}
	\draw[colorp,fill,thick,rot] (\hamx,\y) circle (\thp);
	\draw[colorgap,line width=1.5,rot] ({\hamx-5*\thsq},0) to +(10*\thsq,0);
	\draw [<->,line width=2,rot] ({\hamx+.2},.8) .. controls ({\hamx+.8},.5) and ({\hamx+.8},-.5) .. ({\hamx+.2},-.8);
	\draw[gray] (0,0) circle (1);		
	\draw[cont,domain=40:160] plot ({cos(\x)}, {sin(\x)});
	\draw[cont,domain=200:320] plot ({cos(\x)}, {sin(\x)});
	
	\draw[colorgap,line width=1.5] ({-1.0-5*\thsq},0) to (1+5*\thsq,0);
	
	\foreach \ang in {12,30,168,-12,-30,-168,-180}
	\draw[colorp,fill,thick] ({cos(\ang)}, {sin(\ang)}) circle (\thp);
	\draw [<->,line width=2] ({cos(80)}, {sin(80)-.03}) -- ({cos(80)}, {-sin(80)+.03});
	\end{tikzpicture}
   \caption{\label{fig:gaps}\coloronline
   (a) schematic spectrum of unitary quantum walk with band spectrum (green) and discrete spectrum (red dots) and essential gaps at $\pm1$. The arrow symbolizes the action of the symmetry operators $\ph,\ch$ on quasi-energies. We are interested in the eigenvalues on the symmetry axis indicated, around which we assume an essential gap.
  (b) spectrum of self-adjoint Hamiltonian marked analogously.}
\end{figure}
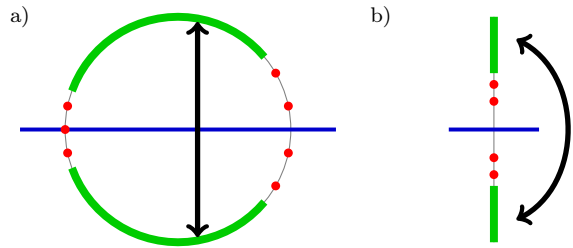

\lettersection{Basic Features}
Let us outline some basic features of our theory. First of all we consider infinite one-dimensional systems only, but make no assumption of translation invariance whatsoever. Lattice sites are labelled by integers $x$, and the variables at that site are given by a finite dimensional Hilbert space $\HH_x$. In the translationally invariant systems all the $\HH_x$ must be the same and equal to the ``coin space'' of the walk, typically two-dimensional. The walk is a unitary operator $W$ on the total Hilbert space $\HH=\bigoplus_{x\in\Ir}\HH_x$, with a finite interaction length $L$. That is, matrix elements between sites further apart than some finite $L$ vanish. The discrete symmetries $\{\ph,\rv,\ch\}$ act on each $\HH_x$ separately and we consider only symmetry types forcing the spectrum of the walk to be symmetric with respect to the real axis (see Fig.~\ref{fig:gaps}). The two real points $\pm1$ in the spectrum of the walk $W$ are therefore special. These points are assumed to be in the gap of the bulk systems. Also it is here that the bulk-edge principle predicts protected bound states. We cover both cases by assuming {\it essential gaps} at $\pm1$, which means that in a small interval around $\pm1$, $W$ has at most finitely many eigenvalues (counting degeneracy). This property is automatically preserved under arbitrary local perturbations \cite{Weyl}.

To any walk in this setting (possibly defined only on a half-infinite lattice) we associate an index $\six(W)$ which takes either integer values or the values $0,1$ depending on the symmetry type considered. It is defined in terms of the actions of the symmetry operations in the eigenspaces at $\pm1$ (see below). In particular, $\abs{\six(W)}$ is a lower bound on the total dimension of the two eigenspaces at $\pm1$. For walks on the doubly infinite lattice, this quantity splits into two contributions
\begin{equation}\label{sixRL}
  \six(W)=\sixR(W)+\sixL(W).
\end{equation}
The term $\sixR(W)$ (resp.\ $\sixL(W)$) only depends on the behavior of $W$ far to the right (resp.\ the left). Hence these are independent of any local perturbations, and the pair $\bigl(\sixL(W),\sixR(W)\bigr)$ is the basic invariant of our theory.

A translation invariant system cannot have any isolated eigenvalues of finite degeneracy, so $\six(W)=0$, and $\sixR(W)$ already contains all classifying information. Now consider the typical scenario of bulk-edge correspondence: We take two translation invariant walks $W_1,W_2$, and a crossover between the two, i.e., a walk $W$ which coincides with $W_1$ far to the left and with $W_2$ far to the right. Then we get at least $\abs{\six(W)}=\abs{\sixR(W_2)+\sixL(W_1)}=\abs{\sixR(W_2)-\sixR(W_1)}$ independent eigenvectors at $\pm1$. In particular, we get at least one if the classifying invariants are different, i.e., $\sixR(W_2)\neq\sixR(W_1)$.
All these statements also hold for the Hamiltonian case.
\lettersection{Symmetry types}
In every model we fix a {\it symmetry type} by specifying which of the symmetries $\ph,\rv,\ch$ are present, and also the signs in $\ph^2=\pm\idty$ and $\rv^2=\pm\idty$. The particle hole symmetry $\ph$ and the time reversal symmetry $\rv$ are always antiunitary \cite{Wigner1,Wigner2}, and the chiral symmetry $\ch$ is unitary. When two such symmetries are present, their product provides the third, and we choose the phase convention making them commute. The three signs in Table~\ref{types} then multiply to $1$.
A Hamiltonian $H$ is said to {\it satisfy the symmetry} if  $\ph H=-H\ph$, $\ch H=-H\ch$, and $\rv H=H\rv$. The corresponding conditions for walks can be read off the conditions which the time evolution operator $W=e^{-itH}$ satisfies in this case, i.e., $\ph W=W \ph$, $\ch W=W\dg\ch$, and $\rv W=W\dg\rv$, if the respective symmetry is present. The symmetries are assumed to act in each cell $\HH_x$ separately. We also assume that the symmetry operators in each cell are {\it balanced} meaning that there is some unitary acting only in the cell, which satisfies the symmetry and has no eigenvalues $\pm1$.

\begin{table}\begin{centering}
\begin{tabular}{r|c||c|c|c||c|c|c}
       &$S$  &$\ph^2$  &$\rv^2$ &$\ch^2$  &$\ig(S)$ & $\six$\\ \hline
    1  &D    &$\idty$   &         &           &$\Ir_2$   & $d\,\mod2$\\[3pt]
    2  &AIII &          &         &$\idty$    &$\Ir$     & $\tr\ch$\\[3pt]
    3  &BDI  &$\idty$   &$\idty$  &$\idty$    &$\Ir$     & $\tr\ch$ \\[3pt]
    4  &CII  &$-\idty$  &$-\idty$ &$\idty$    &$2\Ir$    & $\tr\ch$ \\[3pt]
    5  &DIII &$\idty$   &$-\idty$ &$-\idty$   &$2\Ir_2$  & $ d\,\mod4$
\end{tabular}
 \caption{\label{types}Symmetry types considered in this paper, with their generators and relations for their squares. Absence of an entry in the respective column means that the generator is not part of the type.
 The second column gives the Cartan classification \cite{Altland-Zirnbauer}. The perhaps unusual sign of $\ch^2$ in the last row is due to the convention which makes the three symmetry operators commute.
 $\ig(S)$ is the range of the symmetry index $\six$ of a finite dimensional representation of the symmetry.
 The last column gives an explicit expression for $\six$, where $d$ denotes the dimension of the representation. }
  \label{symtab}
\end{centering}\end{table}

Given the symmetry operators we now consider quantum walks $W$ which (1) satisfy the symmetry and (2) are essentially gapped. These assumptions carry over in an obvious manner to the Hamiltonian case.

\lettersection{Defining the symmetry index}
The main idea for the definition of the symmetry index $\six(W)$ is now suggested by perturbation theory. We want the index to be stable under any small deformation of $W$ which respects the symmetry. As an example, consider the first symmetry type ($\ph$ only). Then, since eigenvalues $\neq\pm1$ occur in pairs, the dimension of the eigenspace at $+1$ can change, but only by multiples of $2$. The same holds for the eigenspace at $-1$ and hence for the total dimension. We thus take $\six(W)=0,1$ as the parity of the combined dimensions of the eigenspaces at $\pm1$.

We can generalize this to all the other symmetry types by observing that the change of the eigenspaces can only be by a subspace, on which the symmetries act in a balanced way as defined in the previous section. Concretely, for the types 2,3,4 in the table, consider an eigenvector $W\phi=\lambda\phi$ with $\im\lambda\neq0$ then $\ch\phi$ is an eigenvector with eigenvalue $\overline{\lambda}$. Therefore, on the subspace spanned by $\phi,\ch\phi,\ph\phi,\rv\phi$, $\ch$ acts like a swap operator, which has zero trace. This is the sort of subspace by which the eigenspaces of $W$ at $\pm1$ may change. Therefore, if $\espace$ denotes the sum of the eigenspaces of $W$ at $\pm1$ we can take $\six(W)=\tr_\espace\ch$ with the trace evaluated just over $\espace$. This ensures that $\six(W)$ is invariant under all small deformations satisfying the symmetry. The technical details of this argument and the appropriate version for the 5.\ symmetry type will be given in \cite{LongVersion}.

\lettersection{The splitting construction}
The second main ingredient of our theory is a way to split a walk into a right and a left block, acting only on cells $x< x_0$ and $x\geq x_0$, respectively. This is done by a unitary operator $V$, which differs from the identity only on finitely many cells around $x_0$, i.e., is a local perturbation. The perturbed walk is then to be of the form $W'=VW=W_L\oplus W_R$, and must satisfy the symmetry. It is crucial that we can choose $V$ to be a {\it gentle perturbation}, i.e., one that is continuously connected to the identity such that the walks connecting $W$ and $W'$ all satisfy the symmetry. We then define $\sixR(W)=\six(W_R)$ and $\sixL(W)=\six(W_L)$. By invariance of $\six(W)$ under small perturbations, $\six(W)=\six(VW)=\six(W_L)+\six(W_R)$, which is \eqref{sixRL}. The core of the decoupling construction stems from \cite[Lemma~4]{OldIndex}, and details will be given in \cite{LongVersion}. We have to exclude walks of the last symmetry type in Table~\ref{symtab} at this point, because we could not establish in full generality the existence of a decoupling which satisfies the symmetry.

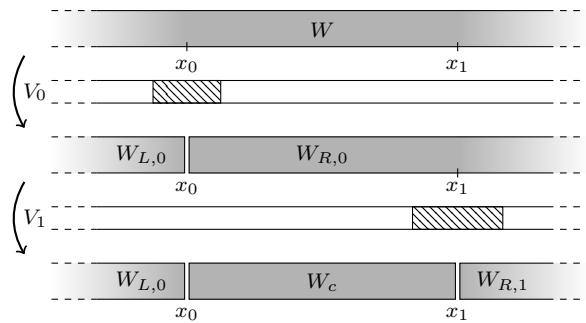
\begin{figure}
\begin{center}
	\begin{tikzpicture}
	[
	scale=1.2,
	font=\footnotesize,
	dsh/.style={dash pattern=on 2.5pt off 2.5pt},
	innertext/.style={fill=white,rounded corners=2pt,inner sep=1pt}
	]
	
	\colorlet{shadecol}{gray!60}
	
	\def\xa{1.5}
	\def\xmax{6.}
	\def\ya{.4}
	\def\yb{.25}
	\def\yu{-.5+\yb/2}
	\def\sp{.025}
	
	\shade [left color=white, right color=shadecol] (0,0) rectangle +(\xa,\ya);
	\shade [right color=white, left color=shadecol] (\xmax,0) rectangle +(-\xa,\ya);
	\draw[color=shadecol,fill] (\xa,0) rectangle +({\xmax-2*\xa},\ya);
	
	\shade [left color=white, right color=shadecol,yshift=-1.4cm] (0,0) rectangle +({\xa-\sp},\ya);
	\shade [right color=white, left color=shadecol,yshift=-1.4cm] (\xmax,0) rectangle +(-\xa,\ya);
	\draw[color=shadecol,fill,yshift=-1.4cm] ({\xa+\sp},0) rectangle +({\xmax-2*\xa},\ya);
	
	\shade [left color=white, right color=shadecol,yshift=-2.8cm] (0,0) rectangle +({\xa-\sp},\ya);
	\shade [right color=white, left color=shadecol,yshift=-2.8cm] (\xmax,0) rectangle +(-{\xa+\sp},\ya);
	\draw[color=shadecol,fill,yshift=-2.8cm] ({\xa+\sp},0) rectangle +({\xmax-2*\xa-2*\sp},\ya);
	
	\draw[dsh] (0,\ya) -- +(0.5,0) (0,0) -- +(.5,0) (\xmax,\ya) -- +(-0.5,0) (\xmax,0) -- +(-.5,0);
	\draw[dsh,yshift=-1.4cm] (0,\ya) -- +(0.5,0) (0,0) -- +(.5,0) (\xmax,\ya) -- +(-0.5,0) (\xmax,0) -- +(-.5,0);
	\draw[dsh,yshift=-2.8cm] (0,\ya) -- +(0.5,0) (0,0) -- +(.5,0) (\xmax,\ya) -- +(-0.5,0) (\xmax,0) -- +(-.5,0);
	
	\draw[dsh] (0,\yu) -- +(0.5,0) ++(0,-\yb) -- +(.5,0) (\xmax,\yu) -- +(-0.5,0) (\xmax,{\yu-\yb}) -- +(-.5,0);
	\draw[dsh,yshift=-1.4cm] (0,\yu) -- +(0.5,0) ++(0,-\yb) -- +(.5,0) (\xmax,\yu) -- +(-0.5,0) (\xmax,{\yu-\yb}) -- +(-.5,0);
	
	\draw (.5,0) -- +({\xmax-1},0)  (.5,\ya) -- +({\xmax-1},0);
	
	\draw (\xa,-.05) -- +(0,.1) +(0,0)  node[below] {$x_0$};
	\draw ({\xmax-\xa},-.05) -- +(0,.1) +(0,0)  node[below] {$x_1$};
	\draw ({\xmax/2},.19) node {$W$};	
	
	\draw (.5,-1) -- ({\xa-\sp},-1) -- ++(0,-\ya) -- (.5,{-1-\ya});
	\draw ({\xmax-.5},-1) -- ({\xa+\sp},-1) -- ++(0,-\ya) -- ({\xmax-.5},{-1-\ya});
	
	\draw (.5,{-2-\ya}) -- ({\xa-\sp},{-2-\ya}) -- ++(0,-\ya) -- (.5,{-2-2*\ya});
	\draw ({\xa+\sp},{-2-\ya}) rectangle +({\xmax-2*\xa-2*\sp},-\ya);
	\draw ({\xmax-.5},{-2-\ya}) -- ({\xmax-\xa+\sp},{-2-\ya}) -- ++(0,-\ya) -- ({\xmax-.5},{-2-2*\ya});
	
	\draw [->,thick] (-.3,-.1) to[bend right] node[pos=0.5,right] {$V_0$}  +(0,-.8);
	\draw [->,thick] (-.3,-1.5) to[bend right] node[pos=0.5,right] {$V_1$}  +(0,-.8);
	
	\draw[yshift=-1.4cm] (\xa,.05) +(0,-.05)  node[below] {$x_0$};
	\draw[yshift=-1.4cm] ({\xmax-\xa},-.05) -- ++(0,.1) +(0,-.05) node[below] {$x_1$};
	\draw[yshift=-1.4cm] (1.,.19) node {$W_{L,0}$};	
	\draw[yshift=-1.4cm] ({\xmax/2},.19) node {$W_{R,0}$};	
	
	\draw[yshift=-2.8cm] (\xa,.05) +(0,-.05)  node[below] {$x_0$};
	\draw[yshift=-2.8cm] ({\xmax-\xa},.05) +(0,-.05)  node[below] {$x_1$};
	\draw[yshift=-2.8cm] ({\xmax/2},.19) node {$W_c$};	
	\draw[yshift=-2.8cm] (1,.19) node {$W_{L,0}$};	
	\draw[yshift=-2.8cm] ({\xmax-1},.19) node {$W_{R,1}$};
	
	\draw (0.5,\yu) -- ({\xmax-.5},\yu) (.5,{\yu-\yb}) -- ({\xmax-.5},{\yu-\yb});
	\draw[yshift=-1.4cm] (0.5,\yu) -- ({\xmax-.5},\yu) (.5,{\yu-\yb}) -- ({\xmax-.5},{\yu-\yb});
	\draw[pattern=north west lines] (\xa,\yu) +(-1.5*\yb,0) rectangle +(1.5*\yb,-\yb) ;
	\draw[pattern=north west lines,yshift=-1.4cm] ({\xmax-\xa},\yu) +(-2*\yb,0) rectangle +(2*\yb,-\yb) ;
	
	\end{tikzpicture}
\end{center}
\caption{Sketch of the decouplings needed to prove the independence of the index $\sixR(W)$ of the cut position. It is assumed that $x_0\ll x_1$ such that the nontrivial supports of $V_0$ and $V_1$ do not overlap.}
\label{fig:twocuts}
\end{figure}

\lettersection{Cut-independence}
The topological stability of the symmetry index hinges on the observation that $\sixR(W)$ (and analogously $\sixL(W)$) is  independent of the cut position. To see this, consider two cut positions $x_0,x_1$ with decoupling unitaries $V_0,V_1$ where ${x_1-x_0}$ is sufficiently large such that the regions where these operators differ from the identity do not overlap. We allow $V_0$ to be an arbitrary decoupling, but choose $V_1$ to be gentle, as we may. Then $V_iW=W_{L,i}\oplus W_{R,i}$, for $i=1,2$. Moreover, the piece $W_{R,0}$ is further decoupled by $V_1$, so that $V_1W_{R,0}=W_C\oplus W_{R,1}$ (see Fig.~\ref{fig:twocuts}) for some symmetric $W_C$ on the cells $x_0,\ldots,x_1-1$. Because $V_1$ is gentle (i.e., contractible to the identity), we
have
\begin{equation}\label{2cut}
    \six(W_{R,0})=\six(W_C)+\six(W_{R,1}).
\end{equation}
We will presently argue that $\six(W_C)=0$. Hence the candidates for $\sixR(W)$, computed for different cut positions coincide, and we can take the cut position $x_1$ as large as we wish.

It remains to consider $\six(W_C)$ for an arbitrary unitary $W_C$ satisfying the symmetry on a finite collection of cells, i.e., on the space $\HH_C=\bigoplus_{x=x_0}^{x_1}\HH_x$. To this end we split this space into the sum $\HH_{\rm es}$ of the eigenspaces of $W_C$ at $\pm1$ and the orthogonal complement $\HH_{\rm es}^\perp$. The symmetry operators are block diagonal in this decomposition and, e.g., $\ch_C=\ch_{\rm es}\oplus\ch_{\rm es}^\perp$. On $\HH_{\rm es}^\perp$ the symmetries act in a balanced way, because $W_C$ acts as a gapped unitary satisfying the symmetry. Hence $\tr\ch_{\rm es}^\perp=0$. Similarly, because on each cell, and hence on $\HH_C$ the symmetry operators are balanced, $\tr\ch_C=0$. Since $\tr\ch_C=\tr\ch_{\rm es}+\tr\ch_{\rm es}^\perp$, we get $\six(W_C)=\tr\ch_{\rm es}=0$. The argument with just particle-hole symmetry is analogous.

\lettersection{Identifying topologically protected edge states}
In the existing literature (see e.g. \cite{Asbo1,Kita}) topologically protected edge states are detected indirectly by starting a joined-up walk at the boundary and letting it run for a few steps. Then one looks for some probability getting trapped near the boundary. This can be followed nicely in a Mathematica demonstration written by Kitagawa.

In order to rigorously pinpoint the eigenvalues we used a general method for reducing the eigenvalue problem for a locally perturbed infinite system to that finite subsystem on which the perturbation happens. The influence of the infinite system is then taken into account by the matrix valued {\it Schur function} of the unperturbed system (which is related to the Green function). Some background on Schur functions can be found in \cite{recurrence,BGVW,QDapproach} and references therein. For any unitary $W$ on $\HH$ and any finite dimensional subspace $\HH_0\subset\HH$ the Schur function $f(z)$ is an analytic function on the open unit disc with values in the operators on $\HH_0$, satisfying $\norm{f(z)}\leq1$. In an essential gap $f$ is also analytic and unitary on the unit circle. The eigenspace of $W$ for eigenvalue $z$, or rather the eigen-subspace in the span of all $W^n\HH_0$ is then canonically isomorphic to the fixed point space of the unitary $zf(z)$. This is ideally suited for perturbations on finite dimensional subspaces, because if $V$ is such a unitary differing from $\idty$ only on $\HH_0$, the Schur function of $VW$ is simply $f_{VW}(z)=f(z)V\dg$ \cite{QDapproach}. Moreover, if $\HH_0$ is chosen consistent with a discrete symmetry, this whole construction is also consistent with the symmetry, so one can immediately read off the eigenspaces and the relevant representations of the symmetry, and hence obtain $\six_\pm(VW)$.

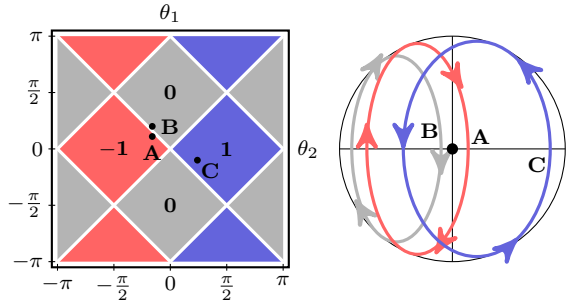
\begin{figure}[h]
\begin{center}
	\tikzset{
		>=stealth',
		big arrow/.style={
			very thick,
			postaction={decorate}
			},
		big arrow reverse/.style={
			very thick,
			decoration={markings, mark=at position 0.17 with {\arrow[thick, scale=2]{>}},
				mark=at position 0.5 with {\arrow[thick,scale=2]{>}},
				mark=at position 0.83 with {\arrow[thick,scale=2]{>}}},
			postaction={decorate}
			},
		right graphic/.style={
			xshift=2.5cm
		}
	}
	
	\begin{tikzpicture}
	[
	scale=1.5,
	font=\footnotesize
		]
		
		\definecolor{pmcol}{RGB}{180,180,180}
		\definecolor{nncol}{RGB}{180,180,180}
		\definecolor{nmcol}{RGB}{100,100,220}
		\definecolor{pncol}{RGB}{255,100,100}
		
		\draw[thick] (-1.05,-1.05)  rectangle +(2.1,2.1);
		
		\fill[nncol] 	(-.5,-.5) -- +(-.5,-.5) -- +(-.5,.5)
								(-.5,-.5) -- ++(.5,.5) -- ++(.5,-.5) -- +(-.5,-.5)
								(.5,-.5) -- +(.5,.5) -- +(.5,-.5);
		\fill[pncol] 	(-.5,-.5) -- +(-.5,-.5) -- +(.5,-.5)
								(-.5,-.5) -- ++(.5,.5) -- ++(-.5,.5) -- +(-.5,-.5)
								(-.5,.5) -- +(.5,.5) -- +(-.5,.5);		
		\fill[pmcol] 	(-.5,.5) -- +(-.5,-.5) -- +(-.5,.5)
								(-.5,.5) -- ++(.5,.5) -- ++(.5,-.5) -- +(-.5,-.5)
								(.5,.5) -- +(.5,.5) -- +(.5,-.5);
		\fill[nmcol] 	(.5,-.5) -- +(-.5,-.5) -- +(.5,-.5)
								(.5,-.5) -- ++(.5,.5) -- ++(-.5,.5) -- +(-.5,-.5)
								(.5,.5) -- +(.5,.5) -- +(-.5,.5);
								
		\draw[white,very thick] (-1,-1) -- (1,1);
		\draw[white,very thick] (1,-1) -- (-1,1);
		\draw[white,very thick] (-1,0) -- (0,1) (0,-1) -- (1,0);
		\draw[white,very thick] (-1,0) -- (0,-1) (0,1) -- (1,0);
		
		\foreach \i in {-1,-.5,0,.5,1}{
			\draw[align=left] (-1.02,{\i}) -- (-1.08,{\i});
			\draw[align=left] ({\i},-1.02) -- ({\i},-1.08);
		}
		
		\draw (-1.05,-1)  node[left,align=left]{ $-\pi$};
		\draw (-1.05,-.5)  node[left,align=left]{ $-\frac{\pi}{2}$};
		\draw (-1.05,0)  node[left,align=left]{ $0$};
		\draw (-1.05,.5)  node[left,align=left]{ $\frac{\pi}{2}$};
		\draw (-1.05,1)  node[left,align=left]{ $\pi$};
		
		\draw (-1,-1.05)  node[below,align=center]{ $-\pi$};
		\draw (-.5,-1.05)  node[below,align=center]{ $-\frac{\pi}{2}$};
		\draw (0,-1.05)  node[below,align=center]{ $0$};
		\draw (.5,-1.05)  node[below,align=center]{ $\frac{\pi}{2}$};
		\draw (1,-1.05)  node[below,align=center]{ $\pi$};
		
		\fill (-0.16, 0.11)  circle (.03) node[below]{$\bf{A}$};
		\fill (-0.16, 0.20)  circle (.03) node[right]{$\bf{B}$};
		\fill (.24,-.1)  circle (.03) node[below right=-0.1cm]{$\bf{C}$};
				
		\draw (0,1.05) node[above,align=center]{ $\theta_1$};
		\draw (1.05,0) node[right,align=left]{ $\theta_2$};
		
		
		\draw (.5,0) node[align=center]{$\bf{1}$};
		\draw (-.5,0) node[align=center]{$\bf{-1}$};
		\draw (0,-.5) node[align=center]{$\bf 0$};
		\draw (0,.5) node[align=center]{$\bf 0$};
		
		
		
		\draw[right graphic] (-1,0) -- (1,0) (0,-1) -- (0,1); 

		\draw[right graphic] (0,0) circle (1);
		
		\draw[big arrow,nncol,right graphic,		
		decoration={markings, mark=at position 0 with {\arrow[thick,scale=2]{<}},
			mark=at position 0.33 with {\arrow[thick,scale=2]{<}},
			mark=at position 0.67 with {\arrow[thick,scale=2]{<}}}
		]	
		(-.496,0) ellipse (.396 and .825);
		\draw[
		big arrow,pncol,right graphic,		
		decoration={markings, mark=at position 0.17 with {\arrow[thick,scale=2]{<}},
			mark=at position 0.5 with {\arrow[thick,scale=2]{<}},
			mark=at position 0.83 with {\arrow[thick,scale=2]{<}}}
		]
			(-.309,0) ellipse (.449 and .935);

		\draw[big arrow reverse,nmcol,right graphic]	
		 (.216,0) ellipse (.651 and .955);

		\fill[right graphic] (0,0)  circle (.05);

		 \draw[right graphic] (.25,.25)  node[below,align=center]{ $\bf{A}$};
		 \draw[right graphic] (-.2,.3)  node[below,align=center]{ $\bf{B}$};
		 \draw[right graphic] (.75,0)  node[below,align=center]{ $\bf{C}$};
		
	\end{tikzpicture}
\end{center}
\caption{\coloronline Interactive version under \explorer.\\ Left: Phase diagram showing the range of $\sixR(W)$ for translation invariant split-step walks with angle parameters $\theta_1,\theta_2$.
The white lines represent gap closures, hence systems not covered in our setting.
Right: Chirally off-diagonal matrix element of walk parameterized by quasi-momentum. The winding number of this curve around the origin is the topological index for translation invariant systems. This bulk index coincides with the symmetry index $\sixR(W)$, which is shown explicitly for the three points A, B, C.
}
\label{fig:harlekin}
\end{figure}

The Schur function of a quantum walk is readily calculated using the CGMV approach to quantum walks \cite{BGVW,CGMV,recurrence,QDapproach}, which proceeds by reducing the unitary evolution to canonical form, given by the so-called CMV matrices \cite{CGMV,CMVoriginal,BGVW,SimonCMV}. Moreover, by using for $V$ the splitting around the origin introduced above one obtains a half-space walk in CGMV form, which coincides with the given, translation invariant one to the the right of the origin. One thus only needs the Schur function relative to the subspace on which $V$ differs from the identity, i.e. at finitely many cells around $x=0$. From this the exact phase diagram (e.g. Fig.~\ref{fig:harlekin} (left)) follows.

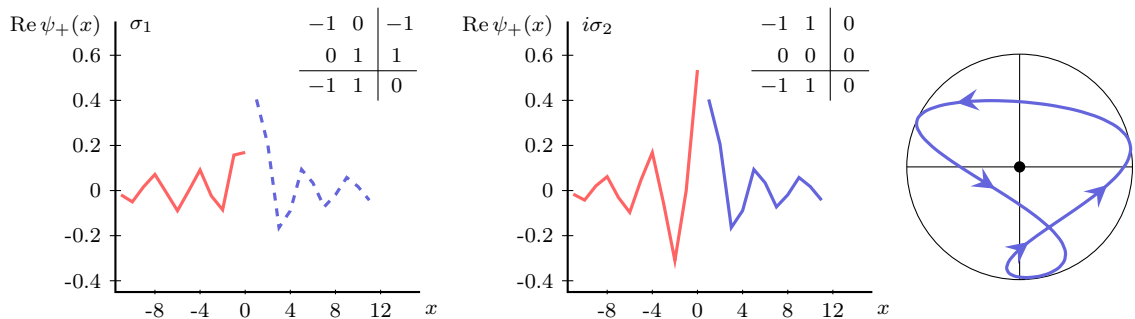
\begin{figure}
	\tikzset{
		fun/.style={
			very thick
			}
	}
	
    \tikzset{
		>=stealth,
		new arrow/.style={
			very thick,
			decoration={markings,
				mark=at position 0.07 with {\arrow[thick, scale=2]{>}},
				mark=at position 0.21 with {\arrow[thick, scale=2]{>}},
				mark=at position 0.55 with {\arrow[thick,scale=2]{>}},
				mark=at position 0.77 with {\arrow[thick,scale=2]{>}}},
			postaction={decorate}
			}
	}	
	\begin{tikzpicture}
	[
	scale=1.5,
	xscale=.1,
	yscale=2,
	font=\footnotesize
	]
	
	\definecolor{chirp}{RGB}{100,100,220}
	\definecolor{chirm}{RGB}{255,100,100}
	
	\draw[thick] (15.5,-.45) -- (-11.5,-.45) -- (-11.5,-.45) -- (-11.5,.75);
	
	\foreach \x in {-8,-4,...,12}
	\draw (\x,-.46) -- (\x,-.44) +(0,-.02) node[anchor=north] {\x};
	\draw (16.5,-.47) node[anchor=north] {$x$};
	
	\foreach \y in {-0.4,-0.2,0,0.2,0.4,0.6}
	\draw (-11,\y) -- (-12,\y) node[anchor=east] {\y};
	\draw (-12,.73) node[anchor=east] {$\mathrm{Re}\,\psi_+(x)$};
	
	\draw (-11,.73) node[anchor=west] {$\sigma_1$};
	
	
	\node[coordinate] (a) at (10,.7) {};
	\matrix(dict)[matrix of nodes,
	nodes={align=center,text width=.3cm},
	] at (10,.6) {
		$-1$ & $0$ & $-1$\\
		$\hphantom{-}0$ & $1$ & $1$\\
		$-1$ & $1$ & $0$\\
	};
	\draw(dict-2-1.south west)--(dict-2-3.south east);
	\draw(dict-1-2.north east)--(dict-3-2.south east);
	
	
	\draw [fun, chirp, dashed] plot coordinates
	{(1, 0.404129) (2, 0.20583) (3, -0.163983) (4, -0.0892989) (5, 0.0934436) (6, 0.034254) (7, -0.0723793) (8, -0.0190517) (9, 0.057842) (10, 0.0175707) (11, -0.042672)}
	;
	
	\draw [fun, chirm] plot coordinates
	{(-11, -0.0191451) (-10, -0.0500888) (-9, 0.0170908) (-8, 0.0716955) (-7, -0.00830534) (-6, -0.0893906) (-5, -0.00061873) (-4, 0.0913879) (-3, -0.0236537) (-2, -0.0852732) (-1, 0.157495) (0, 0.169198)}
	;		
	\end{tikzpicture}
	\begin{tikzpicture}
	[
	scale=1.5,
	xscale=.1,
	yscale=2,
	font=\footnotesize
		]
	
	\definecolor{chirp}{RGB}{100,100,220}
	\definecolor{chirm}{RGB}{255,100,100}
	
	\draw[thick] (15.5,-.45) -- (-11.5,-.45) -- (-11.5,-.45) -- (-11.5,.75);

	\foreach \x in {-8,-4,...,12}
		\draw (\x,-.46) -- (\x,-.44) +(0,-.02) node[anchor=north] {\x};
	\draw (16.5,-.47) node[anchor=north] {$x$};
	
	\foreach \y in {-0.4,-0.2,0,0.2,0.4,0.6}
	\draw (-11,\y) -- (-12,\y) node[anchor=east] {\y};
	\draw (-12,.73) node[anchor=east] {$\mathrm{Re}\,\psi_+(x)$};
	
	\draw (-11,.73) node[anchor=west] {$i\sigma_2$};
	
	
	\node[coordinate] (a) at (10,.7) {};
	\matrix(dict)[matrix of nodes,
	nodes={align=center,text width=.3cm},
	] at (10,.6) {
		$-1$ & $1$ & $0$\\
		$\hphantom{-}0$ & $0$ & $0$\\
		$-1$ & $1$ & $0$\\
	};
	\draw(dict-2-1.south west)--(dict-2-3.south east);
	\draw(dict-1-2.north east)--(dict-3-2.south east);

	
	\draw [fun, chirp] plot coordinates
	{(1, 0.404129) (2, 0.20583) (3, -0.163983) (4, -0.0892989) (5, 0.0934436) (6, 0.034254) (7, -0.0723793) (8, -0.0190517) (9, 0.057842) (10, 0.0175707) (11, -0.042672)}
	;
	
	\draw [fun, chirm] plot coordinates
	{(-11, -0.0159148) (-10, -0.0422743) (-9, 0.0207687) (-8, 0.0616639) (-7, -0.0308917) (-6, -0.096613) (-5, 0.0475891) (-4, 0.168508) (-3, -0.0566716) (-2, -0.310388) (-1, -0.00121322) (0, 0.534344)}
	;		
	\end{tikzpicture}
    \begin{tikzpicture}
	[
    baseline={(0,-2.1)},
	scale=1.5,
	font=\footnotesize
		]
		
        \definecolor{chirp}{RGB}{100,100,220}
		
		\draw (-1,0) -- (1,0) (0,-1) -- (0,1);

		\draw (0,0) circle (1);
		
		\fill (0,0)  circle (.05);
		

		\draw [new arrow, chirp] plot [smooth cycle] coordinates
		{(0.057904, -0.982287) (-0.0308793, -0.972622) (-0.0920003, -0.943865) (-0.114593, -0.896723) (-0.0920895, -0.832359) (-0.0229739, -0.752355) (0.0889811, -0.658684) (0.235009, -0.55365) (0.402151, -0.439841) (0.574497, -0.320058) (0.734727, -0.197251) (0.865819, -0.0744443) (0.95275, 0.0453385) (0.984016, 0.159148) (0.952839, 0.264182) (0.857914, 0.357853) (0.70365, 0.437857) (0.499856, 0.502221) (0.260912, 0.549363) (0.00449632, 0.57812) (-0.25, 0.587785) (-0.483235, 0.57812) (-0.677669, 0.549363) (-0.819153, 0.502221) (-0.898193, 0.437857) (-0.910771, 0.357853) (-0.85865, 0.264182) (-0.749122, 0.159148) (-0.594247, 0.0453385) (-0.409634, -0.0744443) (-0.212904, -0.197251) (-0.021978, -0.320058) (0.146647, -0.439841) (0.279456, -0.55365) (0.367171, -0.658684) (0.405559, -0.752355) (0.395746, -0.832359) (0.343985, -0.896723) (0.260912, -0.943865) (0.160367, -0.972622) (0.057904, -0.982287)}
		;

		
	\end{tikzpicture}
  \caption{Left and Center: Real parts of eigenfunctions of the decoupled walk $W$. The red color encodes chirality $-1$,
  	whereas blue indicates $+1$. A solid line is used when the corresponding eigenvalue of $W$ is $1$, where the dashed line indicates $-1$.
  	The decoupling matrix at $x=0$ is chosen as both, $A_0$ and $B_0$ being the reflection matrix $\sigma_1$ (left) contrasted to the decoupling being realized by both, $A_0$ and $B_0$ being the gentle decoupling $i\sigma_2$. Clearly visible in both cases is the localization on either the left or the right side of the cut position. The insets reflect Table~\ref{invtab} and clearly show that whereas both, $\sixL(W)$ as well as $\sixR(W)$ are invariant under the decoupling chosen, the homotopy invariants $\six_-(W)$ and $\six_+(W)$ are not invariant under non-gentle perturbations, thus confirming the abstract results above.\\
  Right: Typical winding of chirally off-diagonal matrix element of \eqref{eq:wfs} leading to $\sixR(W)=+1$.}
  \label{fig:fourstep}
\end{figure}

\lettersection{Local vs.\ gentle perturbations}
As Fig.~\ref{fig:gaps} shows one difference between the unitary case considered here and the the Hamiltonian case is that we have {\it two} gaps. It is suggestive to try to associate invariants with each of these gaps \cite{Asbo2}, since the basic perturbation argument showing $\six(W)$ to be constant under symmetry preserving deformations applies to both gaps separately. So let us introduce $\six_+(W)$ as the symmetry index of the symmetries acting in the eigenspace of $W$ for eigenvalue $+1$, and similarly $\six_-(W)$ for the $-1$-eigenspace. Combined with a splitting $VW=W_L\oplus W_R$ we get the four numbers in the upper left of Table~\ref{invtab}.
Summing the two columns of this $2\times2$-matrix yields the invariants we have already defined. They are insensitive to any local modification of the walk, and in particular independent of how the cut is made. They are also invariant under arbitrary norm-small perturbations \cite{LongVersion}, although we have not shown that here. Indeed, the most straightforward attempt to prove that actually fails, i.e., it is not possible \cite{LongVersion} to find for every continuous family of walks $t\mapsto W_t$ gentle decouplings $V_t$ which also depend continuously on $t$. On the other hand, if the decoupling is chosen to be gentle, we have that $W_L\oplus W_R$ is a continuous deformation of $W$, and so each row adds up to $\six_\pm(W)$. The gentleness of the decoupling is a crucial condition here: There are also non-gentle, but local decouplings, and for these the sum rule does not hold. Finally, the individual numbers $\six_\pm(W_{L,R})$ have no useful invariance properties at all: they are neither invariant under local nor gentle perturbations, and generally depend on the position of the decoupling cut.
An example of the indices $\six_+(W)$ and $\six_-(W)$ depending on the type of the cut is given in Figure~\ref{fig:fourstep}.

\begin{table}[h]\begin{centering}\def\schpace{\hbox to 12pt{}}
\begin{tabular}{cc|c}
       $\six_+(W_L)$\schpace  &$\six_+(W_R)$\schpace  &\strut\ $\six_+(W)$  \\
       $\six_-(W_L)$\schpace  &$\six_-(W_R)$\schpace  &\strut\ $\six_-(W)$  \\\hline
       $\sixL(W)$\schpace     &$\sixR(W)$\schpace  &\strut\ $\six(W)$ \vbox to 12pt{}
\end{tabular}
 \caption{Table of four potential invariants from the splitting construction.
   The sum in each row gives the marginal shown, provided the decoupling is chosen to be gentle. The column sums are always the invariants discussed in this paper. Individual entries have no invariance properties, but the row sums are invariant under all gentle perturbations. }
  \label{invtab}
\end{centering}\end{table}

To summarize, we have two invariants $\bigl(\sixR(W),\sixL(W)\bigr)$ for all perturbations, but if we only discuss gentle perturbations, there is a third invariant, namely $\six_+(W)$, i.e., the sum of the first row in Table~\ref{invtab}. The second row would give the same information, since $\six_-(W)=\sixR(W)+\sixL(W)-\six_+(W)$.
\lettersection{Example: The split-step walk revisited}
Introduced in \cite{Kita}, this example is also treated in \cite{Kita2,Asbo1,Asbo2,Asbo4}, and many other papers. It has symmetry type 3, see Table~\ref{symtab}. The coin space is two-dimensional, with $\ph$ the complex conjugation in position space. Its name derives from the use of two separate shift operations, $S_{\uparrow}$, the right shift of the spin-up vectors and $S_{\downarrow}$, the left shift of the spin-down vectors. Choosing the local basis in each cell $\HH_x$ appropriately (sometimes called ``choosing a different time-frame'' \cite{Asbo2}) the split-step walk takes the form
\begin{equation}\label{eq:wss}
  W=BS_{\downarrow}AS_{\uparrow}B,
\end{equation}
where $A=\bigoplus_x A_x$ and $B=\bigoplus_x B_x$ are unitary operators acting sitewise, and each $A_x$ , $B_x$ is an admissible unitary operator acting on $\HH_x$. In the basis chosen the chiral symmetry takes the standard form $\ch=\bigoplus_x \sigma_1$ and $\ph$, if applicable, is given by complex conjugation. In the form of \eqref{eq:wss} it is straightforward to see that admissibility of the $A_x$, $B_x$ suffices for the admissibility of $W$. To see this, note that admissibility for $\ph$ depends solely on the reality of the entries of $W$ and hence follows by choosing $A_x=R(\theta_2)$ and $A_x=R(\theta_1/2)$ to be real rotations since $S_\uparrow$ and $S_\downarrow$ are real anyhow. For the chiral symmetry $\ch$, we use the indentities
\begin{equation}
  \ch S_\downarrow=S_\uparrow^*\ch,\qquad\ch S_\uparrow=S_\downarrow^*\ch.
\end{equation}
to prove by straightforward computation that hermiticity of $W\ch$ follows directly from the hermiticity of $B\ch$ and $A\ch$. The phase diagram for the index $\sixR(W)$ is shown in Fig.~\ref{fig:harlekin}. 


This approach to split-step walks allows one to easily generate examples with phase diagrams of arbitrary dimension. The symmetric form suggests a straightforward extension to a split-step walk with three-dimensional phase diagrams by considering
\begin{equation}\label{eq:wfs}
  W=CS_{\uparrow}BS_{\uparrow}AS_{\downarrow}BS_{\downarrow}C.
\end{equation}
The admissibility of this walk follows on the same lines as for \eqref{eq:wss}. To generate an example of symmetry type 2 but explicitly violates particle-hole symmetry $\ph$ the rotation matrices $C$ were garnished by $C\mapsto C\left(\begin{smallmatrix}\hphantom{-}i&\hphantom{-}1\\-1&-i\end{smallmatrix}\right)/\sqrt2$. The  chirally off-diagonal matrix element as well as the eigenfunctions are plotted in~Figure \ref{fig:fourstep}. The specific angles of the rotations $A,B,C$ lead to a winding number $\sixR(W)=+1$.

\lettersection{Summary and Outlook}

We have established a theory of topological invariants for one-dimensional quantum walks with discrete symmetries. The known classification of translation invariant phases is reproduced, and in addition we get a full classification of non-translation invariant systems stable under arbitrary symmetric local perturbations, thus providing a rigorous theory of bulk-edge correspondence in this setting. The technical details will be presented in a future publication \cite{LongVersion}.

We believe that some variant of this theory will also apply to the fifth symmetry type in Table~\ref{types}, which so far is not covered. Moreover, we envisage extensions to higher spatial dimensions and the periodic table \cite{kitaevPeriodic} of symmetric systems and edge modes, at least when the translation symmetry parallel to the interface between two phases is not broken.

\lettersection{Acknowledgements}
We thank Tobias J. Osborne, Andrea Alberti, and J{\'a}nos Asb{\'o}th for a critical reading of the manuscript.

C. Cedzich, C. Stahl and R. F. Werner acknowledge support from the ERC grant DQSIM and the European project SIQS.

F. A. Gr\"unbaum was partially  supported by the Applied Math. Sciences subprogram of the Office of Energy Research, USDOE, under Contract DE-AC03-76SF00098, and by AFOSR grant FA95501210087 through a subcontract to Carnegie Mellon University.

The work of L. Vel\'azquez is partially supported by the research project MTM2011-28952-C02-01 and MTM2014-53963-P from the Ministry of Science and Innovation of Spain and the European Regional Development Fund (ERDF), and by Project E-64 of Diputaci\'on General de Arag\'on (Spain).

A. H. Werner acknowledges support from the ERC grant TAQ.

\ifarxiv
	\bibliography{tphbib}
\else
	\bibliography{tphbib}
\fi 

\end{document}